\documentclass[amsmath,amssymb,amsbsy,prl,nofootinbib,twocolumn,tightenlines,superscriptaddress,floatfix]{revtex4-2}
\usepackage{graphicx}
\usepackage{bm}
\usepackage{amsmath}
\usepackage{amsfonts}
\usepackage{amssymb}
\usepackage[normalem]{ulem}
\usepackage[caption=false]{subfig}
\usepackage{verbatim}

\usepackage[dvipsnames]{xcolor}

\definecolor{refblue}{RGB}{20,80,130}

\usepackage[
    colorlinks=true,
    citecolor=refblue,
    urlcolor=refblue,
    linkcolor=refblue
]{hyperref}

\newcommand{\vel}{\boldsymbol{\mathrm{v}}}

\newcommand{\icts}{International Centre for Theoretical Sciences, Tata Institute of Fundamental Research, Bangalore 560089, India}
\newcommand{\iith}{Department of Physics, Indian Institute of Technology (IIT), Hyderabad, Kandi Sangareddy Telangana, 502285, India}
\newcommand{\iitk}{Department of Mechanical Engineering, Indian Institute of Technology Kanpur, Kanpur 208016, India}

\begin{document}

\title{A Reynolds-like Criterion for the Onset of Universal Active Turbulence}
\author{Kunal Kumar$^\clubsuit$} 
\email{kunal.kumar@icts.res.in}
\affiliation{\icts}
\author{Sandip Sahoo$^\clubsuit$}
\email{sandipsahoo09902@gmail.com}
\affiliation{\icts}
\author{Kirti Kashyap} 
\email{kirtikashyap26@gamil.com}
\affiliation{\iith}
\author{Amal Manoharan}
\email{amalmanoharan1994@gmail.com}
\affiliation{\icts}
\author{Siddhartha Mukherjee}
\email{smukherjee@iitk.ac.in}
\affiliation{\iitk}
\author{Anupam Gupta} 
\email{agupta@phy.iith.ac.in}
\affiliation{\iith}
\author{Samriddhi Sankar Ray}%
\email{samriddhisankarray@gmail.com}
\affiliation{\icts}

\begin{abstract}
Beyond a critical Toner--Tu drive, bacterial turbulence exhibits universal
statistics, including a parameter-independent energy spectrum
$E(k)\sim k^{-3/2}$. We show that the same universal state can be reached
not only by increasing activity, but also by strengthening nonlinear
advection, weakening nonlinear saturation, or tuning the Swift--Hohenberg
instability. Motivated by the competition between these processes, we
identify the independent dimensionless groups governing the dynamics and
construct a Reynolds-like control parameter from them. A single critical
value, calibrated from the known activity-controlled transition, predicts
the onset of universality along the other parameter routes and collapses
them onto a common threshold. These results identify a dynamical organizing
parameter for the onset of universal active turbulence and provide a
Reynolds-like measure for the Toner--Tu--Swift--Hohenberg universality
class.
\end{abstract}

\maketitle
\def\thefootnote{$\clubsuit$}\footnotetext{These authors contributed equally to this work}\def\thefootnote{\arabic{footnote}}

Dense suspensions of self-propelled bacteria, microtubule--motor mixtures,
active nematics and living cells spontaneously generate chaotic flows
characterized by continuously created and destroyed vortices
\cite{mendelson1999organized,dombrowski2004self,
creppy2015turbulence,sanchez2012spontaneous,blanch2018turbulent}.
Although these systems operate at negligible Reynolds number, where inertia
is unimportant, their dynamics can nevertheless resemble those of inertial
turbulence and are therefore collectively termed \emph{active turbulence}.
A broad class of active flows is described by nonlinear continuum theories,
including Toner--Tu, Swift--Hohenberg and active-nematic hydrodynamics
\cite{marchetti2013hydrodynamics,ramaswamy2010mechanics,
dunkel2013fluid,wensink2012meso,dunkel2013minimal,
slomka2015generalized,giomi2015geometry,
thampi2014vorticity,blanch2017hydrodynamic,
ramaswamy2016activity,saintillan2008instabilities,
saintillan2014theory}. These theories generate self-sustained turbulent
dynamics through the competition of active driving, nonlinear transport,
saturation and pattern-forming instabilities, rather than through inertial
instabilities. This raises a fundamental question: \emph{what determines
the onset of universal active turbulence?}

Universality in active matter has recently emerged as a defining feature of
these nonequilibrium flows. Alert \textit{et al.}
\cite{alert2020universal} identified universal large-scale behaviour in
active nematics, while Mukherjee \textit{et al.}
\cite{mukherjee2023intermittency} showed that bacterial turbulence undergoes
a transition beyond a critical Toner--Tu activity
\cite{toner1995long}, above which its statistical properties become
parameter independent \cite{kiran2025onset}. In particular, the energy spectrum approaches the
universal form $E(k)\sim k^{-3/2}$, in close agreement with  experimental observations \cite{Liu_2012}. The existence of this transition
suggests that universality is not tied to a particular microscopic
parameter, but emerges when the competing processes governing the flow
reach an appropriate dynamical balance.

This question is nontrivial because the continuum description contains
several parameters that control physically distinct aspects of the dynamics.
In the Toner--Tu--Swift--Hohenberg (TTSH) model, active driving, nonlinear
saturation and advection coexist with parameters that set the intrinsic
Swift--Hohenberg length and time scales. Experimentally, bacterial
concentration, swimming speed, active stress, confinement and the mechanical
properties of the surrounding medium can modify several of these quantities
simultaneously
\cite{dombrowski2004self,cisneros2011dynamics,
sokolov2007concentration,sokolov2012physical,
ryan2013correlation,peng2021imaging,lemma2019statistical,kumar2018tunable,
aranson2022bacterial}. A criterion based on any individual coefficient would
therefore not capture the different routes through which a system can
approach the universal state. If the onset is independent of the route
through parameter space, it should instead be organized by dimensionless
combinations of the competing dynamical processes, analogous in spirit to
the Reynolds number in inertial turbulence. Unlike the inertial case,
however, it is not obvious how active driving, nonlinear transport,
saturation and pattern formation should combine.

Here we identify the dimensionless combinations governing the TTSH dynamics
and construct from them a Reynolds-like control parameter for the onset of
universal active turbulence. We show that the universal state can be reached
through distinct routes: by increasing active driving, strengthening
nonlinear advection, weakening nonlinear saturation, or tuning the
Swift--Hohenberg instability. Although these routes act on different terms
of the governing equation, their transition boundaries are described by the
same dimensionless criterion. A single critical value of
$Re_{\rm act}$, calibrated from the known activity-controlled transition,
then predicts the onset of universality along the other parameter routes.
This result identifies $Re_{\rm act}$ as a dynamical organizing parameter
for the onset of universal active turbulence and provides a Reynolds-like
measure for the TTSH universality class.

\begin{figure*}
   \centering
    \includegraphics[width=1\textwidth]{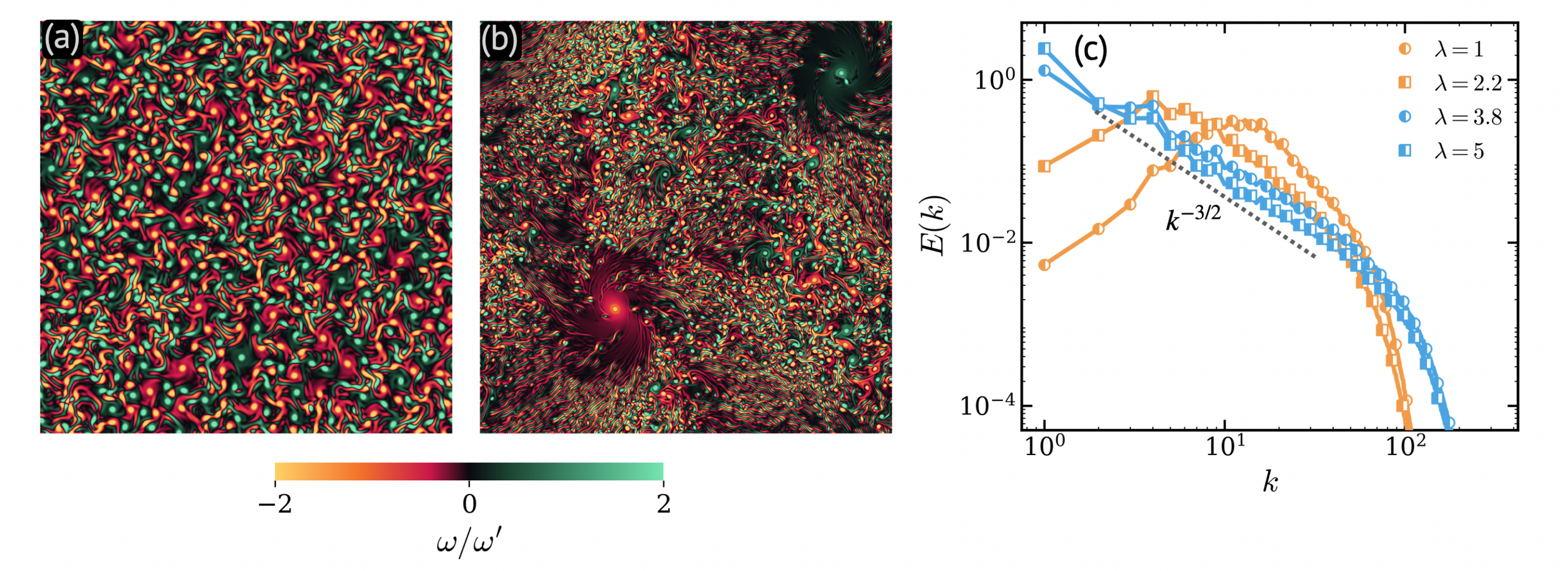}
    \caption{\textbf{Stronger nonlinear advection produces the flow structures
and spectral scaling associated with universal active turbulence.}
Vorticity field $\omega/\omega^\prime$ at fixed $\alpha=-6$ for
(a) $\lambda=1$ and (b) $\lambda=5$. Increasing $\lambda$ transforms a
dense field of small vortices into elongated streaks and larger spiral
vortices, reproducing the characteristic structures previously observed at
strong Toner--Tu activity~\cite{mukherjee2021anomalous}. (c) Energy spectra
for increasing $\lambda$, showing the transition from non-universal spectra
at weak nonlinear advection to the universal scaling
$E(k)\sim k^{-3/2}$ at strong nonlinear advection.}
\label{fig:vorticity}
\end{figure*}

To identify what controls the onset of universality, we now examine the dynamical behavior of the incompressible TTSH equation governing the bacterial
turbulence~\cite{wensink2012meso,dunkel2013minimal,alert2022active,Sahoo_2026},
\begin{equation}
\partial_t \vel +\lambda \vel\cdot\nabla\vel = -\nabla p
-\Gamma_0\nabla^2\vel -\Gamma_2\nabla^4\vel
-\alpha\vel -\beta|\vel|^2\vel ,
\label{eq:ttsh}
\end{equation}
where the pressure $p$ enforces incompressibility,
$\nabla\cdot\vel=0$. The term $\lambda_1\nabla|\vel|^2$
that originally appears in the general TTSH model~\cite{wensink2012meso} is a pure
gradient and has been absorbed into $p$. Throughout we take
$\Gamma_0,\Gamma_2>0$ and $\alpha<0$; the latter is the active regime, in
which $-\alpha\vel$ is a growth term. 

The equation contains three distinct physical
mechanisms. The Toner--Tu terms $(\alpha,\beta)$ provide active driving
and nonlinear saturation, with their balance setting the characteristic
velocity scale
$v_\alpha=\sqrt{\frac{|\alpha|}{\beta}}$
of the homogeneous polar state, which exists only for
$\alpha<0$~\cite{toner1995long}. The advective coupling $\lambda$
controls nonlinear transport. Although
$\lambda\vel\cdot\nabla\vel$ has the form of an inertial term, it does not
arise from fluid inertia, which is negligible here, but from
self-propulsion and active stresses: $\lambda=1$ corresponds to passive
advection, and the microscopic theory gives $\lambda=1-S$, where $S<0$ is
an active-stress parameter for pusher-type swimmers, so that
$\lambda>1$~\cite{wensink2012meso}. Finally, the Swift--Hohenberg terms
$(\Gamma_0,\Gamma_2)$ select a most unstable mode at
$k_m = \sqrt{\frac{\Gamma_0}{2\Gamma_2}}$ and thereby introduce an intrinsic spatial
scale~\cite{cross1993pattern}.

The Swift--Hohenberg coefficients define an intrinsic length
$\ell_\Gamma=\sqrt{\Gamma_2/\Gamma_0}$ and time
$\tau_\Gamma=\Gamma_2/\Gamma_0^2$, together with the coefficient-balance
wavenumber $k_\Gamma=\ell_\Gamma^{-1}=\sqrt{\Gamma_0/\Gamma_2}$, at which
the second- and fourth-order Swift--Hohenberg contributions in
Eq.~\eqref{eq:ttsh} are equal in magnitude and cancel.
The rate $\tau_\Gamma^{-1}=\Gamma_0^2/\Gamma_2$ is therefore the size of
either of these two competing terms at $k_\Gamma$, not a net growth rate.

\begin{figure*}[htbp]
     \centering
    \includegraphics[width=\textwidth]{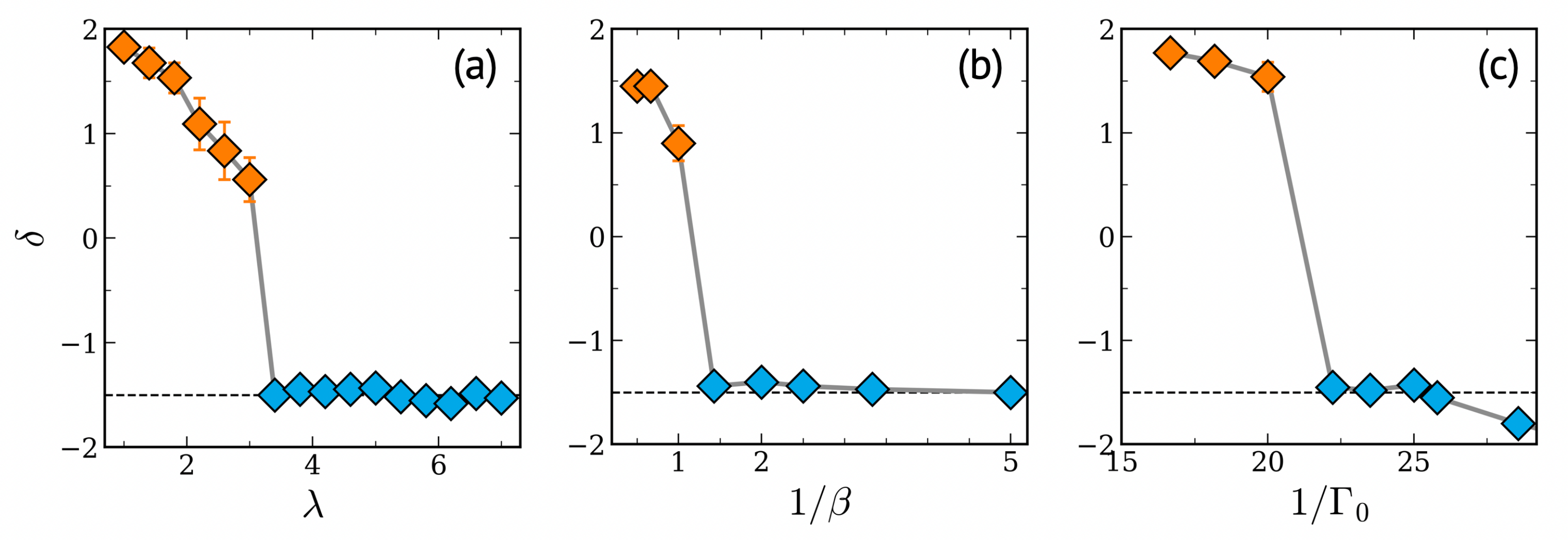}
\caption{\textbf{Universal active turbulence is reached through distinct
routes in TTSH parameter space.}
Energy spectra for increasing (a) advective coupling $\lambda$,
(b) nonlinear saturation coefficient $\beta$, and (c) inverse
Swift--Hohenberg parameter $1/\Gamma_0$, with
$\alpha=-6$ and $\Gamma_2=9\times10^{-5}$ fixed. In each case, the spectral
exponent approaches the common universal value $\delta_c=-3/2$ beyond a
sharp transition. The corresponding critical values are
$\lambda_c\approx3.4$, $\beta_c\approx0.7$, and
$\Gamma_{0,c}\approx0.045$.}
	\label{fig:energy_spectra}
\end{figure*}

Following Ref.~\cite{mukherjee2023intermittency}, we diagnose universality
in Eq.~\eqref{eq:ttsh} through the energy-spectrum exponent,
$E(k)\sim k^\delta$, which distinguishes the universal from the
non-universal regimes. The exponent is obtained from a
least-squares fit over the range $k_{\min}\le k\le k_{\max}$
of time-averaged spectra in the statistically steady state, and the
transition point of a parameter scan is defined as the value at which
$\delta$ crosses $-3/2$.
Unless otherwise stated, the parameters are fixed at $\alpha=-6$,
$\lambda=3.5$, $\beta=0.5$, $\Gamma_0=0.045$ and $\Gamma_2=9\times10^{-5}$,
and one of them is varied at a time. We first consider the effect of
nonlinear advection on the flow morphology. Figure~\ref{fig:vorticity}
compares two simulations at fixed $\alpha=-6$ for (a) $\lambda=1$ and (b)
$\lambda=5$. Increasing $\lambda$ transforms a dense field of small
vortices into elongated streaks and fewer, larger spiral vortices, resembling the structural transition typically observed with increasing activity~\cite{Kashyap_2026}. Similar streaky
structures were reported by Mukherjee
\textit{et al.}~\cite{mukherjee2021anomalous} under strong Toner--Tu drive,
where they were associated with anomalous L\'evy-walk transport. Their
emergence here through enhanced nonlinear advection shows that the same
large-scale flow organization can arise through a different microscopic
route.

The corresponding energy spectra are shown in
Fig.~\ref{fig:vorticity}(c) for increasing $\lambda$. At weak nonlinear
advection the spectrum is non-universal, whereas for sufficiently strong
advection it develops the parameter-independent scaling
$E(k)\sim k^{-3/2}$. The crossover between these regimes provides a
spectral signature of the transition to universal active turbulence.
Thus, enhanced nonlinear advection not only reorganizes the flow
morphology but also drives the system towards the universal spectral state
previously identified at strong Toner--Tu
activity~\cite{mukherjee2023intermittency}.

Before extending this to other parameters, we note that
$\lambda$ and $\beta$ are not independent in Eq.~\eqref{eq:ttsh}. The
rescaling $\tilde{\mathbf v}=\lambda\mathbf v$, $\tilde p=\lambda p$
removes $\lambda$ from the advective term and replaces the cubic
coefficient by $\beta_{\mathrm{eff}}=\beta/\lambda^2$, leaving all other
terms unchanged. Increasing $\lambda$ at fixed $\beta$ is therefore
exactly equivalent to decreasing $\beta$ at fixed $\lambda$; the two
enter only through the combination $\beta/\lambda^2$, i.e.\ through the
advective velocity scale $\lambda v_\alpha=\lambda\sqrt{|\alpha|/\beta}$.
Since the rescaling multiplies $E(k)$ by $\lambda^2$ at every $k$, it
leaves $\delta$ unchanged, which is one reason the spectral exponent is a
convenient diagnostic. A $\lambda$ scan and a $\beta$ scan are thus the
same scan, and their transition points must satisfy
$\beta_c/\lambda^2=\beta/\lambda_c^2$ exactly. We nevertheless perform
both, as a consistency check on the numerics and on the definition of the
transition point.

We next ask whether the same transition can be reached by modifying other
parameters of the TTSH dynamics. Figure~\ref{fig:energy_spectra} shows
$\delta$ as a function of $\lambda$, $\beta$, and $\Gamma_0$, with the
remaining parameters at their default values. In all three cases, the
exponent changes over a narrow parameter range and approaches the same
asymptotic value, $\delta_c=-3/2$. Below the transition the exponent
depends on the parameter being varied, indicating non-universal
statistics, whereas above it the spectra converge to the same scaling.
The transition points are $\lambda_c\approx3.4$, $\beta_c\approx0.7$, and
$\Gamma_{0,c}\approx0.045$.
The $\lambda$ and $\beta$ values are consistent with the
exact relation $\beta_c/\lambda^2=\beta/\lambda_c^2$ to within the
resolution of the scans.
Variation of the Toner--Tu drive at the default values of the other
parameters gives a transition at $\alpha_c\approx-5.5$, consistent with
Ref.~\cite{mukherjee2023intermittency}. The deviation from the
asymptotic value at large $1/\Gamma_0$ in
Fig.~\ref{fig:energy_spectra}(c) signals the onset of a condensate, which
bounds the universal regime from above and is not considered further
here.

Together, these results show that universal active turbulence can be
reached through distinct microscopic routes: by strengthening nonlinear
advection or, equivalently, weakening nonlinear saturation; by tuning the
Swift--Hohenberg coefficient $\Gamma_0$; or by increasing the Toner--Tu
drive. This motivates the search for a dimensionless combination of the
competing dynamical processes that organizes the transition across
parameter space.

\begin{figure*}
	\includegraphics[width=1.0\linewidth]{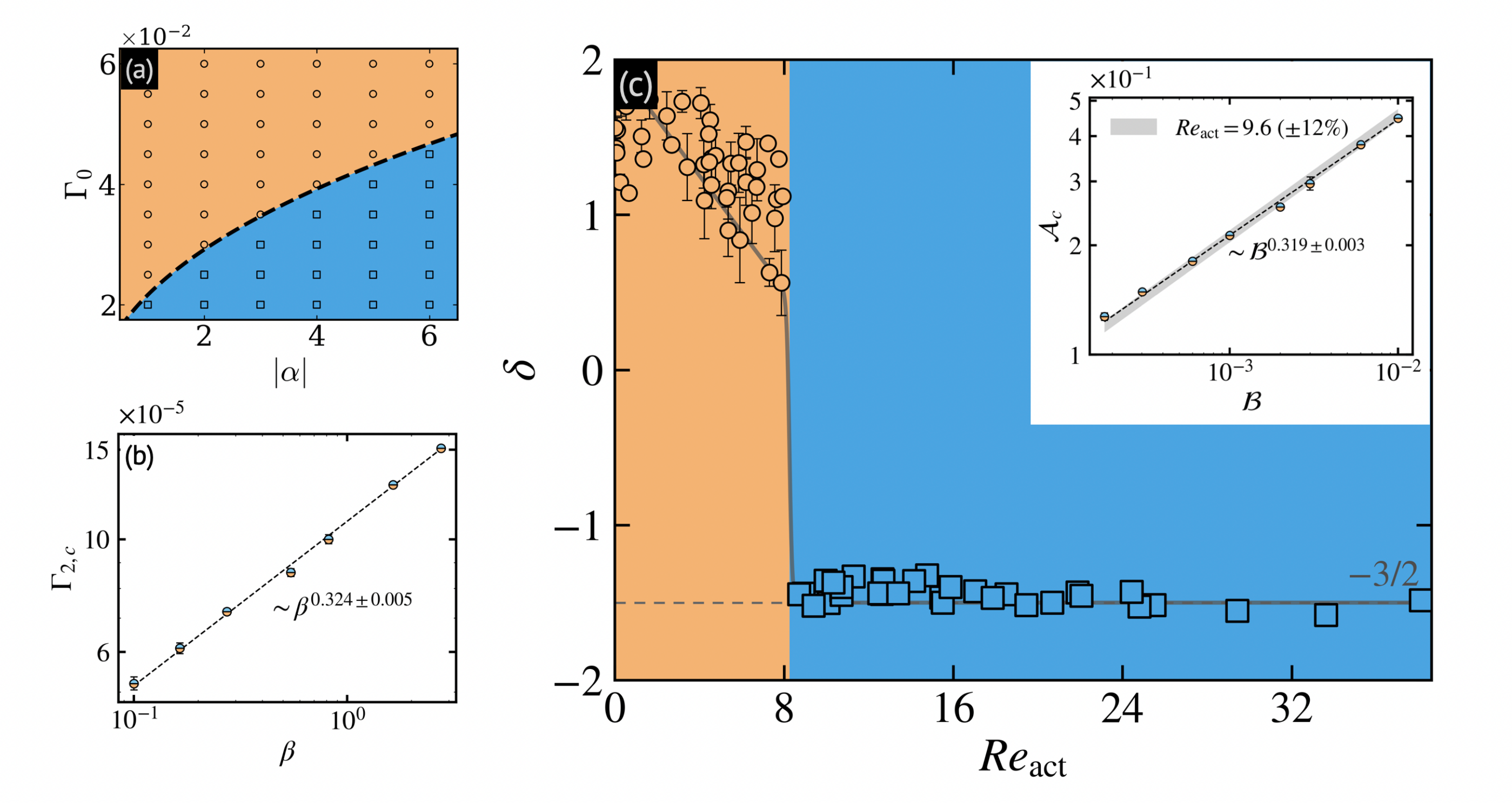}
	\caption{\textbf{A single dimensionless parameter predicts the onset of
universal active turbulence across parameter space.}
(a) Phase diagram in the $|\alpha|$--$\Gamma_0$ plane,
with the remaining parameters held fixed at the values used in Fig.~\ref{fig:energy_spectra}.
The dashed curve indicates the predicted transition from non-universal (above) to universal (below) active turbulence.
(b) Variation of $\Gamma_2$ and $\beta$ at the default values of the other parameters, shown with error bars, reveals that $\Gamma_{2,c} \propto \beta^{1/3}$, or equivalently, ${\mathcal{A}}_c \propto \mathcal{B}^{1/3}$ at the onset of universality. This scaling motivates the definition $Re_{\rm act}\equiv\frac{\mathcal A^3}{\mathcal B}$, with $Re_{\rm act}=Re_{\rm act,c}$ marking the transition.
(c) Plot of the spectral exponent $\delta$ against $Re_{\rm act}$ from the five parameter scans, showing a collapse around a critical value of $Re_{\rm act}$, and demonstrating its robustness. In the inset, we extensively show the scaling dependence of ${\mathcal A}_c$ on ${\mathcal B}$, which agrees with the estimate in (b) within error bars. The mean critical value, $Re_{\rm act,c} = 9.6$, with an uncertainty of approximately $\pm 12 \%$, further supports the estimate obtained from the five parameter scans.
}
    \label{fig:phase_diag}
\end{figure*}

The distinct routes to universal active turbulence identified above suggest that the transition is governed not by any individual TTSH coefficient but by a combination of the competing dynamical processes. We first identify the independent dimensionless groups of the TTSH equation and then determine empirically which combination organizes the transition.

Using the intrinsic Swift--Hohenberg scales $\ell_\Gamma$ and $\tau_\Gamma$ and the velocity scale $U=\ell_\Gamma/(\lambda\tau_\Gamma)$, we write $\mathbf v=U\mathbf u$ and obtain the dimensionless TTSH equation
\begin{equation}
\partial_t\mathbf u+\mathbf u\cdot\nabla\mathbf u
=-\nabla q+\mathcal A\mathbf u-\mathcal B|\mathbf u|^2\mathbf u
-\nabla^2\mathbf u-\nabla^4\mathbf u,
\label{eq:ttsh_dimless}
\end{equation}
where $\mathcal A=\frac{|\alpha|\Gamma_2}{\Gamma_0^2}=|\alpha|\tau_\Gamma$ and $\mathcal B=\frac{\beta\Gamma_0}{\lambda^2}$ are the two dimensionless groups through which the dynamics depend on the five TTSH coefficients in an unbounded domain.
In a periodic box, a third group, $L/\ell_\Gamma$, enters as well; throughout we assume this is sufficiently large that finite-domain effects do not affect the transition. The condensate observed at small $\Gamma_0$, where $\ell_\Gamma$ grows, is the regime in which this assumption fails and is excluded below. The onset of universality must therefore be represented by a transition curve $\mathcal A_c(\mathcal B)$. Nondimensionalization fixes the variables in which this transition is described, but does not determine the shape of the curve.

The two groups have a direct interpretation as ratios of rates. $\mathcal A$ compares the active growth rate $|\alpha|$ with the intrinsic Swift--Hohenberg rate $\tau_\Gamma^{-1}$. Meanwhile, $\mathcal A/\mathcal B = (\lambda v_\alpha k_\Gamma\tau_\Gamma)^2$ is the squared nonlinear-transport rate at the intrinsic wavenumber measured in units of the Swift--Hohenberg rate. Thus $\mathcal A$ measures active driving and $\mathcal A/\mathcal B$ measures nonlinear transport relative to the intrinsic pattern-forming dynamics. For the default parameters, $\sqrt{\mathcal A/\mathcal B}\simeq12.1$, so nonlinear transport is already more than an order of magnitude faster than the intrinsic Swift--Hohenberg rate. The universal state is therefore characterized by activity that is marginal to pattern selection, but strongly dominant nonlinear transport.

We now determine $\mathcal A_c(\mathcal B)$ empirically. The phase diagram between $\alpha$ and $\Gamma_0$ in Fig.~\ref{fig:phase_diag}(a) provides a qualitative picture of the non-universal to universal transition marked by the broken contour, while the other three parameters are kept fixed at their default values. To estimate the transition quantitatively, we vary $\beta$, which only changes $\mathcal{B}$, and note down the corresponding transition values for $\Gamma_2$, which controls $\mathcal{A}$ only. Figure~\ref{fig:phase_diag}(b) thus reveals that $\Gamma_2 \propto \beta^{1/3}$, or equivalently $\mathcal{A}_c \propto \mathcal{B}^{1/3}$. This scaling is further verified in Fig.~\ref{fig:phase_diag}(c) inset, where transition values of $\mathcal{A}$ (i.e., $\mathcal{A}_c$) is directly plotted against $\mathcal{B}$. The scaling exponents in both these tests match within the error bars.

Adopting the empirical scaling dependence between $\mathcal{A}$ and $\mathcal{B}$ at the onset of transition, we define
\begin{equation}
Re_{\rm act}\equiv\frac{\mathcal A^3}{\mathcal B}
=\frac{\lambda^2|\alpha|^3\Gamma_2^3}{\beta \, \Gamma_0^7},
\label{eq:ReTTSH}
\end{equation}
so that the transition occurs on the level set $Re_{\rm act}=Re_{\rm act,c}$. Any monotone function of $\mathcal A^3/\mathcal B$ would represent the same transition boundary; the normalization above is chosen so that $Re_{\rm act}=\mathcal A^2(\mathcal A/\mathcal B)$, i.e. the squared activity multiplied by the squared transport ratio.

Unlike the Reynolds number, $Re_{\rm act}$ is not derived as a ratio of two rates. Its structure is fixed empirically by the transition geometry. The parameter is therefore a Reynolds-like organizing parameter for the TTSH dynamics: nondimensionalization identifies the relevant parameter space, the measured transition boundary gives the cubic power law, and the resulting single-variable criterion is then tested against the independently varied $\Gamma_0$ scans and the collapse of all transition data.

The criterion~\eqref{eq:ReTTSH} contains one number to be fixed, the critical value $Re_{\rm act,c}$. The robustness of this number is verified when all transitions, along any routes, coincide around a single value with minimal error bar. Figure~\ref{fig:phase_diag}(c) exactly provides this test. The spectral exponents from
the five parameter ($\lambda$, $\alpha$, $\beta$, $\Gamma_0$ and $\Gamma_2$) scans collapse around a critical value of $Re_{\rm act}$. The uncertainty is set by the resolution of the transition scans.

We can also obtain a similar transitional $Re_{\rm act}$ when working with $\mathcal A$ and $\mathcal B$. In Fig.~\ref{fig:phase_diag} inset, while exploring the transitional value of $\mathcal A$ over an extended range of $\mathcal B$, we find an average  $Re_{\rm act,c} = 9.6$, with an error of about $\pm 12 \%$.
This suggests that,
over the parameter range explored, the onset of universal active turbulence is
organized by a single Reynolds-like combination of the TTSH parameters.

We have shown that, within the Toner--Tu--Swift--Hohenberg model and over
the parameter range explored here, the onset of universal active
turbulence is organized by a single dimensionless combination of its
hydrodynamic parameters. Although universality can be reached by changing
physically distinct parameters---active driving, nonlinear transport,
saturation, or the Swift--Hohenberg coefficients---the transition occurs
near the same critical value of $Re_{\rm act}$. Thus, $Re_{\rm act}$ should be regarded as an organizing parameter for this hydrodynamic universality class rather than as a universal
Reynolds number for active fluids in general. The
structure of the result is worth stating precisely. Nondimensionalization
reduces the five TTSH coefficients to two groups, $\mathcal A$ and
$\mathcal B$, so that some transition curve $\mathcal A_c(\mathcal B)$
must exist. The empirical findings suggest that this curve is a power law,
$\mathcal A_c\propto\mathcal B^{1/3}$, over a range of around two decades in $\mathcal B$.
This exponent is measured, not derived, and we have
no argument for its value; understanding it is the main open theoretical
question raised by this work.

Whether analogous combinations emerge in other active continuum theories, including active
nematics and more general active suspensions, remains an unexplored area.
Such comparisons may reveal which aspects of the criterion are specific to
the Swift--Hohenberg length and time scales, and which reflect the more general
principles of active nonequilibrium dynamics.

The criterion may also provide a practical means of comparing systems in
which several microscopic parameters vary simultaneously. Changes in
bacterial concentration, swimming speed, active stress, or confinement can
modify multiple effective hydrodynamic coefficients, making direct
comparison between experiments difficult. Expressing the distance to the
transition in terms of $Re_{\rm act}$ could instead provide a common
dimensionless measure across such systems, provided that the corresponding
effective TTSH parameters can be determined.

More broadly, our results show that active turbulence, despite operating
in a regime where inertia is negligible, can possess an organizing
principle analogous to the dimensionless control parameters of classical
fluid mechanics. The relevant competition is not between inertia and
viscosity, but between active driving, nonlinear transport, and intrinsic
pattern-forming dynamics. Identifying analogous dimensionless criteria in
other nonequilibrium active systems may provide a route toward a broader
classification of universal active flows.

K. Kumar thanks C. Rajarshi for helpful discussions.
 SSR acknowledges the Indo–French Centre for the Promotion of Advanced Scientific Research (IFCPAR/CEFIPRA, project
no. 6704-1) for support. This research was supported in
part by the International Centre for Theoretical Sciences
(ICTS) for the program — 11th Indian Statistical Physics
Community Meeting (code: ICTS/11thISPCM2026/04).
The simulations were performed on the ICTS clusters
Mario, Tetris, and Contra. AM, K. Kumar, SS, and SSR acknowledge the
support of the Department of Atomic Energy, Government
of India, under project no.RTI4019 and RTI4013. SM acknowledges the Govt. of India grant ANRF/ECRG/2024/002467/ENS and the IITK Initiation Grant IITK/ME/2024316. K. Kashyap acknowledges the TCS Foundation for financial support through the Research Fellowship Program. A. G. acknowledges SERB-DST (India) Projects MTR/2022/000232, CRG/2023/007056-G for financial support.

\bibliographystyle{apsrev4-2}
\bibliography{Refs}
\end{document}